\documentclass[12pt,preprint]{aastex}

\usepackage{psfig}

\begin{document}

\title{FORMATION AND EVOLUTION OF MICROQUASAR GRS~1915+105}

 \author{KRZYSZTOF BELCZYNSKI\altaffilmark{1,2,3}, TOMASZ BULIK\altaffilmark{2}}

 \affil{
     $^{1}$ Northwestern University, Dept. of Physics \& Astronomy,
       2145 Sheridan Rd., Evanston, IL 60208\\
     $^{2}$ Nicolaus Copernicus Astronomical Center,
       Bartycka 18, 00-716 Warszawa, Poland;\\   
     $^{3}$ Lindheimer Postdoctoral Fellow\\  
     belczynski@northwestern.edu, bulik@camk.edu.pl}

 \begin{abstract} 
Using the {\em StarTrack} population synthesis code we analyze the 
possible formation channels of the Galactic microquasar GRS~1915+105
harboring the most massive known stellar black hole.
We find that we are able to reproduce the evolution and the current 
state of GRS~1915+105 in a model with lowered wind mass loss rates 
in massive stars.
We conclude that the existence of GRS~1915+105 implies that stellar   
winds in massive stars are a factor of two weaker, but still within
observational bounds, than used in stellar models.
We discuss the mass transfer initiated by the massive primary in 
binaries which are the possible precursors of systems harboring 
massive black holes.
We show that such mass transfer evolution is neither clear case B 
nor C, but instead is initiated during the core helium burning phase 
of the massive donor star, the progenitor of the black hole.
We also argue that massive black holes do not receive any kicks at 
their formation and are formed through direct collapse of progenitor 
star without accompanying supernova explosion.  
 \end{abstract}


\section{INTRODUCTION}

The binary GRS~1915+105 is a member of the group called microquasars.
It is a stellar size analogue of quasars, containing an accreting black hole
and ejecting two jets at relativistic speeds (Mirabel \& Rodriguez 1998).
A number of groups have attempted to find the 
basic properties of the system: Morgan, Remillard \& Greiner (1997) 
and Greiner, Morgan \& Remillard (1998) used the RXTE data to constrain the mass 
of the black hole (BH) and inferred the limits between $10$ and $30\ M_\odot$.
Recently Greiner, Cuby, \& McCaughrean (2001) used the infrared observations to 
obtain the mass function. They were also able to identify and characterize the 
donor. 
They find that the system consists of a $14\pm 4\,M_\odot$ BH, 
with a $1.2\pm 0.2\ M_\odot$  K-M III type giant filling its Roche lobe, 
at an orbital separation of $108 \pm 4 R_\odot$ (period of 33.5 days).

The formation of such system poses a problem for the 
standard stellar evolutionary scenarios (Greiner et al. 2001). 
The mass of the BH is larger than the BH masses
measured so far. 
In the framework  of stellar evolution it was argued (Wellstein \& Langer
1999) that massive stars are not able to produce such massive BHs.
The high wind mass loss rates (or the mass transfer event) remove entire H-rich 
envelopes of massive stars, which become massive Wolf-Rayet stars with 
yet more enhanced mass loss rates.
Therefore, at the time of core collapse/SN explosion these massive stars 
are reduced to a fraction of their initial mass, and they can not be
responsible for formation of massive BHs.
In order to avoid this problem, it was suggested that wind mass loss 
rates were overestimated (e.g. Nugis \& Lamers 2000).
It was also proposed that probably the mass transfer (MT) phase reveling bare helium core
happens very late in the evolution of a star (case C MT), and thus remaining
lifetime of Wolf-Rayet phase is very short and wind mass loss is not that
significant (Brown, Lee \& Tauris 2001; Kalogera 2001).
However, Nelemans \& van den Heuvel (2001) performed calculations with
reduced helium mass loss rates and allowed for case C MT scenario, and 
they have found that they are still not able to reproduce the formation 
of high mass BHs.
This was due to the fact, that the case C MT is attainable only for 
primaries initially less massive than 19-25 $M_\odot$, and thus not able 
to form high mass BHs. 
Nelemans \& van den Heuvel (2001) suggested that  helium stars end their
lives with higher masses than it is currently believed (i.e., either
their lifetimes are shorter or the mass loss rates should be further
reduced).

Recent development of population synthesis codes may offer a solution for 
formation of high mass BHs in binaries. 
Belczynski, Kalogera \& Bulik (2002, hereinafter BKB02) have shown that 
using detailed hydrodynamical calculation of core collapse (Fryer 1999) 
and allowing for a direct BH formation there is a possibility of forming 
BHs of $\sim 10 M_\odot$ for a wide range of initial stellar masses.
They have also demonstrated that reduction of lowered helium star wind 
mass loss rates by another factor of 2 (which is still allowed by the 
observations) may increase maximum BH mass formed out of a single star 
up to $\sim 15 M_\odot$, while reduction of all wind mass loss rates 
(both for H-rich and He-rich stellar phases) increases the maximum BH 
mass to $\sim 19 M_\odot$. 
Belczynski, Bulik, Kluzniak (2002) have calculated that the mass 
of BH may be maximally increased during binary interactions by 
$\sim 4 M_\odot$. 
The highest mass BHs appear because of MT in a binary, 
yet this is not likely in case of GRS~1915+105.
The BH of GRS~1915+105 is in a binary with a low mass star, and 
therefore BH mass could not have been increased significantly.
Currently the mass ratio of the system components is $q \approx 0.1$. 
However, the progenitor of the BH must have been even more massive
and the initial mass ratio must have been even smaller, $q_{ini} < 0.05$.
To appear at the present orbital separation the system must have gone
through a common envelope (CE) stage.
It is not clear when the CE event took place and what type of MT 
(case B, C) was involved (Brown, Lee, \& Tauris\ 2001, 
Nelemans \& van den Heuvel 2001)

In this paper we present the investigation of the possible
evolutionary paths leading to the presently observed parameters of 
GRS~1915+105. We use the {\em StarTrack} population synthesis 
code (BKB02) to search for possible evolutionary paths that may lead 
to formation of such a binary. We describe  the model and present the
results in \S~2, while  \S~3 contains the discussion and summary.

\section{CALCULATIONS AND RESULTS}

In our calculations we use the {\em StarTrack} population synthesis code
described in detail in BKB02. 
In our standard evolutionary scenario, we use modified Hurley et al. (2000)
formulae to evolve single stars along with their proposed wind mass loss
rates.
Different types of binary interactions are taken into account, and in
particular we follow the evolution of dynamically stable and unstable 
MT phases. To model dynamically stable events we follow
prescription proposed by Podsiadlowski et al. (1992) and for unstable 
phases we invoke standard CE prescription of Webbink (1984).
We form compact objects in supernova (SN) explosion/core collapse events,
following the results of hydrodynamical calculations of Fryer (1999).
Neutron stars are formed in full flagged SN explosions and receive 
natal kicks form the Cordes \& Chernoff (1998) distribution; intermediate
mass BHs are formed through partial fall back and in attenuated
SN explosions receiving smaller kicks than neutron stars (NSs); while most 
massive BHs are formed through direct collapse of immediate progenitor 
with no accompanying SN explosion and they do not received any kicks 
at their formation.

\subsection{Parameter Space--Evolutionary Considerations}

We start by   examining
the properties of the binaries resembling the current state of 
GRS~1915+105, i.e. containing a BH accreting from a low mass giant.
In order to trace the evolution of GRS~1915+105 like system in more detail
we require that, under the assumption of conservative MT orbit
evolution, the period of the binary will at some point 
reach the current value of $33.5$\,days, and that then 
the BH mass is $=10 M_\odot<M_{\rm BH}<18 M_\odot$, and the red giant with 
mass of $1.0 M_\odot<M_{\rm RG}<1.4 M_\odot$.
As during conservative MT from lower mass object period increases, this
leaves only systems  with periods shorter than 
the observed one as possible progenitors of GRS~1915+105.
The systems could have gone through two MT episodes:
first when the more massive star (currently the BH) was evolving,
and we are witnessing the second one now.

Under the assumption that the second MT phase began already when the 
donor was a
red giant, we calculate the  maximum mass loss rate from giant donor in binary
with a massive BH.
Maximum mass loss increases with the lifetime of a giant (shorter for
higher masses) and with core mass of a giant at the onset of MT (bigger for
higher masses).
Using formulae presented by Ritter (1999) and stellar models implemented in
{\em StarTrack} we find that only red giants with masses $\leq 1.75 M_\odot$
have a chance to become less massive than $1.4 M_\odot$.
A $2 M_\odot$ star may lose only $0.25 M_\odot$ during its red giant
lifetime ($t_{\rm rg} \sim 23$ Myrs), while $1.6 M_\odot$ star 
($t_{\rm rg} \sim 110$ Myrs) may lose up to $0.4 M_\odot$ to a 
$10-18 M_\odot$ BH companion.  
Under the above assumptions, we  obtain $\sim 1.75 M_\odot$ upper limit 
on the mass of secondary in GRS~1915+105, which suggests that only a small 
amount of mass was transferred in the system, and that the observed BH mass 
is close to the one, at which the BH was formed.

There is also a possibility, that the current MT onto BH began when
the secondary was still unevolved, i.e. on the Main Sequence (MS).
However the chances of such configuration are much smaller than for
a giant donor.
The reason is that low-mass MS stars expand only slightly 
(factors of a few) and to form a BH system with MS Roche lobe overflowing 
donor would require that it  starts from the orbital periods in a very 
narrow range.
On the other hand, for red giants the
 radius expansion is significantly larger (factors of
tens to several tens), so the systems from much wider range of orbital periods 
may lead to Roche lobe overflow.
Moreover, if the MT began when the secondary was still on MS, it is not clear
that such a system would survive the evolution through the secondary rapid 
expansion across the Hertzsprung Gap and 
later form MT system with a giant donor, 
resembling the one observed in GRS~1915+105. 

Thus in our simulations we are looking for systems with the orbital 
period shorter than $33.5$\,days, with a BH more massive than $10 M_\odot$ 
and a giant with the mass lower then $1.75 M_\odot$ and effective temperature 
in the range of $4000$--$4750$\,K corresponding to spectral type range: K4--K0. 
We evolved  binaries with the massive primaries ($20-100\,M_\odot$), and 
the low-mass secondaries ($0.5-3\,M_\odot$). 
We first ran a standard model simulation (see BKB02) with $5 \times 10^6$ 
binaries and obtained no systems satisfying the above defined criteria. 
For such low initial mass ratios the masses of the BHs in binaries 
barely reach up to $10 M_\odot$. We then ran a set of simulations with 
modified binary evolution parameters:
a model  with all stellar winds decreased by two (model G1 of BKB02),
a model with increased CE efficiency $\alpha_{CE}\lambda=3.0$ (model E4),
and a model where BH receive the same magnitude kicks
as the NSs do (model S1),
and a model in which BH receive no kicks at birth regardless of their 
mass (model S2). Only one model (model G1) was successful,
and we obtained $435$ GRS~1915+105 like systems
after evolving $5 \times 10^6$ primordial binaries. 

In most of the cases shown the  progenitor system evolves through a CE phase, 
when the primary expands during its evolution on the giant
branch. This tightens the orbit to periods resembling the one observed today.
However, although infrequently, the system may start from a much wider 
orbit, and thus avoid a CE phase (the primary never expands enough to fill 
its Roche lobe). 
Then at the time of the BH formation a correctly placed natal kick 
puts the BH in a tight orbit with its companion, and  that was how several 
systems obtained a period resembling the  one observed today in GRS~1915+105.    
With the proper period and BH in place, the system then just awaits till the
low-mass secondary evolves off the MS and becomes a red giant.
Expansion along the red giant branch eventually brings the secondary to fill
its Roche lobe, and the second MT episode starts. During this phase the mass
is transferred   conservatively from the secondary to the BH,
causing the increase of orbital period. Somewhere along this phase, which may
last hundreds of Myrs for low-mass giants, the system  resembles the
currently observed state of  GRS~1915+105.

\subsection{Example of Formation}

\begin{figure*}[t]
\centerline{\psfig{file=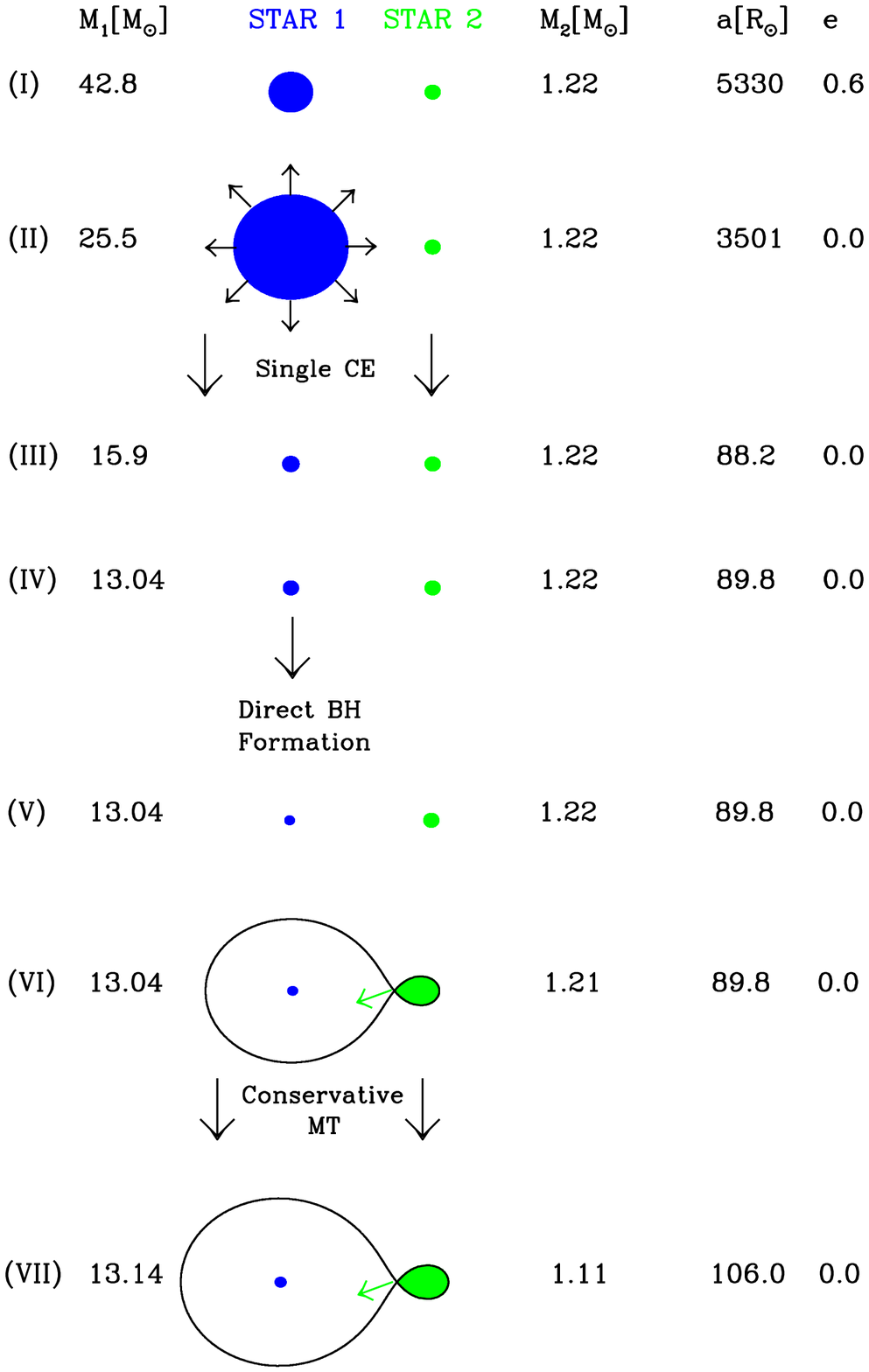,width=0.75\textwidth}}
\caption
{Example of evolutionary channel leading to formation of GRS~1915+105.
Starting with two ZAMS stars (stage I) the systems evolves (after 6 Gyrs) 
to form a Roche Lobe filling red giant (K3 III) on the close orbit 
($P=33.5$ days) around a massive BH (stage VII).
Details of evolution are described in \S\,2.2.
}
\label{MQevol}
\end{figure*}

In Figure~\ref{MQevol} we show an example of binary evolution leading 
to formation of GRS~1915+105.
In order to reproduce the high BH mass we were forced to decrease   
all mass loss rates used in our standard model  by a factor of two.
We start the evolution with two Population I (Z=0.02) Zero Age Main Sequence 
(ZAMS) stars, a $42.8 M_\odot$\ primary and a $1.22 M_\odot$\ secondary, 
in a wide 
(semi-major axis $a=5330 R_\odot$) and eccentric orbit (eccentricity
$e=0.6$). 
In about $4.6$\,Myrs the primary evolves off the MS 
and becomes a
giant, increasing its size significantly.
When the radius of the primary 
becomes comparable to the periastron distance tidal
circularization becomes efficient and the orbit circularizes.
We treat circularization with the conservation of binary angular momentum,
and thus $a$ decreases.
After further expansion the primary fills its Roche Lobe, and although it has
lost almost 17 $M_\odot$\ in the wind, it is still much more massive than
its low-mass MS companion.
Due to the extreme mass ratio, the following MT is dynamically 
unstable and leads to CE phase (stage II). 
The envelope of the primary is ejected from the system at
 the cost of the binary orbital
energy, and the binary orbit shrinks by more than an order of magnitude.
Since the CE phase was very short (evolution on dynamical timescale) the
secondary is almost unaffected, however the primary after the loss of its
entire envelope becomes a massive helium star (stage III).
The helium star primary evolves through consecutive nuclear burning phases,
loosing $\sim 3 M_\odot$\ in the heavy Wolf-Rayet star type wind (slight
expansion of the orbit, stage IV), until it finally collapses to form a BH.
The immediate BH progenitor is a helium star, 
which due to its high mass implodes 
directly to form a  BH without any associated SN explosion. 
There is neither mass loss nor orbital parameter (a,e) change (stage
V). A small mass (stellar binding energy) loss in this phase would introduce
a small eccentricity which will not affect significantly the further evolution
of the system.

The binary now harboring the 
 massive BH and the low-mass MS secondary remains unchanged on
a circular $26$\,day orbit for over 5 Gyrs, 
the time needed for the secondary to evolve 
off the MS.
The secondary on its way up the red giant branch fills its Roche Lobe, and the
second  MT begins (stage VI).
The donor is much less massive than the BH companion and the MT is dynamically 
stable, proceeding on nuclear timescale of expansion of the secondary.
The conservative MT rate may be approximated with a simple analytical 
formulae (e.g. Ritter 1999).
Vilhu (2002) estimated the current 
MT rate for GRS~1915+105 to be $\dot{M}_d=(1.5 \pm 0.5) 
\times 10^{-8} M_\odot yr^{-1}$.
Since the predicted MT rate may be overestimated due to X-ray heating of 
the donor (Vilhu 2002) we use $\dot{M}_d=1.0\times 10^{-8} M_\odot yr^{-1}$  
to evolve the system through the MT phase.
Starting from the initial period of $P_i=26^d.1$ the MT continues through 
10.5 Myrs until the  orbit expands to match the observed period of GRS
1915+105 of $P_f=33^d.5$ (stage VII). 
We have assumed that all the material transferred from the donor ($0.1
M_\odot$) was accreted onto BH.
This  assumption (mass conservation) does not hold exactly in case of
microquasars as at least a fraction  
 of the transferred material must be lost in the  jets.
Had we assumed non-conservative evolution during MT, we would have 
obtained a similar result, although we would expect the MT to last diffrent 
period of time (depending on the specific angular momentum of ejecta) in 
order for the binary to reach the observed period of GRS~1915+105.

\section{DISCUSSION AND CONCLUSIONS}

\begin{figure*}[t]
\centerline{\psfig{file=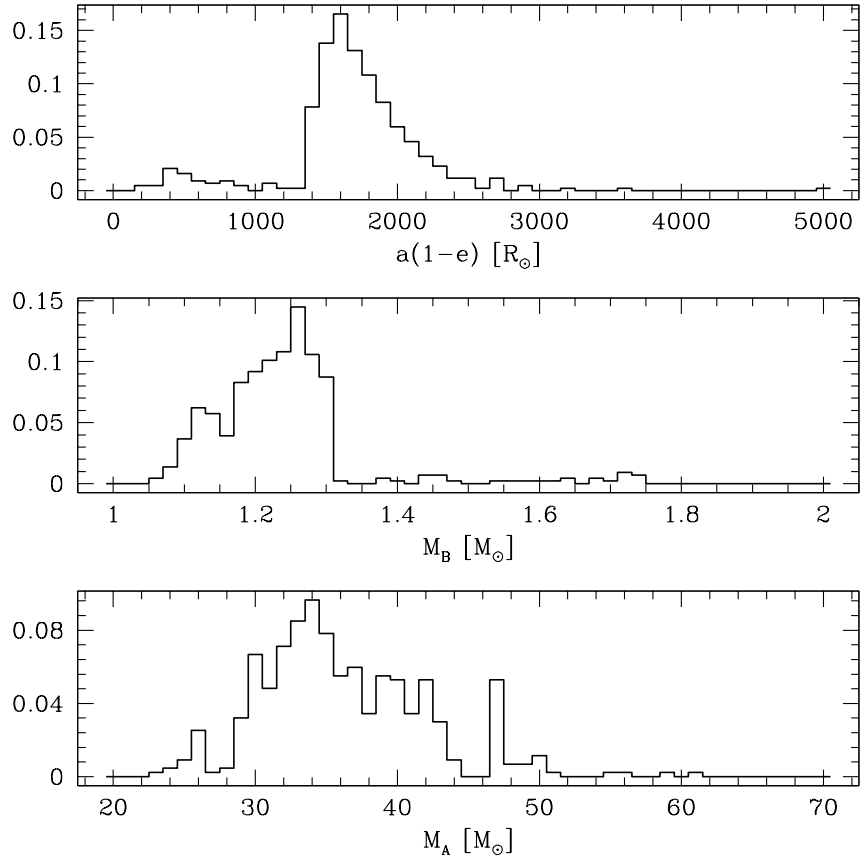,width=0.75\textwidth}}
\caption
{The distributions of the initial parameters: the periastron distance 
(top panel), the mass of the secondary (middle panel), and the mass of
the primary (bottom panel) of binaries that lead to formation of a 
GRS~1915+105 like system. Each distribution is normalized to unity.
}
\label{initial}
\end{figure*}

We have modeled the evolutionary paths
leading to formation of  binaries
with a massive BH  accreting from a low mass
giant  using the {\em StarTrack} 
population synthesis code. 
The initial binary separations span a wide range
however the initial periastron distance  $a(1-e)$
is constrained in the range between $1400$ and 
$2400 R_\odot$ (see Figure~\ref{initial}). 
This is due to the fact  that we require the period of 
the system to become smaller than the one observed today in GRS~1915+105.  
Wider systems either have too long resulting periods 
or never enter a CE phase. 
The tighter systems begin the first MT
just when they leave the MS
and their cores are not massive enough to produce
later the observed high mass BH.
The distribution of the initial masses of the primary, shown in
Figure~\ref{initial}, is quite wide. It starts rising steeply above 
$\sim 27 M_\odot$, where direct BH formation starts (for lowered 
wind mass loss rates; see Fig.1 in BKB02).
It means that {\em i)} masses of BH are high and that {\em ii)} BH do not 
receive natal kicks, which would easily disrupt many potential 
progenitors. 
The distribution peaks at around $\sim 34 M_\odot$ and then declines 
roughly as dictated by combination of our assumed initial mass 
function and the initial-final mass relation.
In the model G1 where potential GRS~1915+105 progenitors appear, we find that 
even stars with $M_{zams} > 40 M_\odot$  may overfill 
their Roche lobe and initiate 
MT.
Such massive stars are believed to lose mass at the rate sufficiently high
that in response the orbit expands faster than the star increase its size,
hence MT is not expected.
However, we should keep in mind that in our model we have decreased wind mass 
loss rates, and therefore the orbital expansion is much slower.
Most of the secondary  masses lie between $1.1-1.3 M_\odot$, 
implying that only small amount of material  has been accreted 
onto BH.

We find that the large mass of the BH  
can be easily explained provided that 
the wind mass loss rates are decreased by a factor 
of two. The wind mass loss rates in massive stars 
have been discussed by Nelemans \& van den Heuvel (2001).
In their Figure 2 describing the mass loss
rates for Wolf-Rayet stars there is quite a significant scatter
in the observational points, and the only point for an $\approx 40\ M_\odot$ 
star lies a factor of a few below all  lines corresponding
to different approximations used in the literature.
Thus we conjecture that the mass loss rates for 
massive Wolf-Rayet stars above $\approx 20\ M\odot$ could be approximately
a factor of two lower than it would seem based on the extrapolation 
of the results for lower mass stars.

We note that the first MT (CE phase) from the massive 
primary to the low mass secondary proceeds in most cases
not by the clear case B or case C evolution.
This fact was already noted in the literature (e.g. Dewi et al. 2002), 
however it is still not widely acknowledged. 
Massive stars ($M_{\rm zams} \geq 14 M_\odot$) begin core helium burning
(CHeB) shortly after leaving MS, and in most cases they avoid an extended 
phase of H-shell burning (first giant branch).
On the H-R diagram these stars move to the  up and right,
increasing monotonically their temperature and luminosity.
In contrast to intermediate and low-mass stars, their radii increase
during the CHeB phase.
As the CHeB phase is relatively long-lived, many of these stars overfill 
their Roche lobes if harbored in close binaries, driving a MT during 
CHeB.
Standard MT terminology devised for low-mass and intermediate 
mass stars, which contract during CHeB, and thus can not drive MT 
during this phase, is therefore not applicable for very massive 
stars.
To avoid the confusion, we denote here the 
MT driven by the star during 
its CHeB as late case B MT.

Formation of the most massive BH in close binary systems needs to
be connected with this late case B evolution, as this is the stage 
when the star has the biggest core when it is stripped of its envelope.
Additionally, one may form a massive BH in close binaries from initially
very wide systems, which avoided MT but then during the BH formation a
 precisely
placed natal kick may tighten the binary.
However, the possibility of such an event is very small, and moreover it
is not yet clear if most massive BHs receive any kicks at all.
In fact, Nelemans, Tauris \& van den Heuvel (1999) found that spatial
velocities of binaries harboring BH, are easily explained by the symmetrical 
SN mass ejection and without any significant natal kick accompanying BH 
formation. 
We consider the low spatial velocity of GRS~1915+105 ($\gamma = 
-3 \pm 10$\ km s$^{-1}$), as the first observational evidence 
that the most massive BH are formed without any kicks imparted on the 
system, and thus no accompanying mass ejection nor SN explosion.
This is just a reasonable extrapolation of the findings of Nelemans et al. 
(1999) which is consistent with theoretical modeling of SN/core collapse
events by Fryer (1999); NS are formed in asymmetric SN explosions, while
intermediate massive BH are formed through partial fall back accompanied by 
almost symmetric SN and receive either a small or no kick at all, 
while the most
massive BH are formed in direct collapse with no SN nor natal kick.  

We conclude that existence of a massive BH in GRS~1915+105 implies that
the currently used wind mass loss rates, both for H- and He-rich stars,
are possibly overestimated by factor of $\sim 2$ for very massive stars.
Moreover, our models reproduce the observed properties of GRS~1915+105, 
only if we allow for direct BH formation, with no accompanying SN explosion.
Our results also suggest that the directly formed BH do not receive any 
natal kicks. 
If we agree on the possibility of the direct collapse of a massive star at
the end of its nuclear lifetime, and also if we allow for change of wind  
mass loss rates within observational bounds, we find that evolution and   
formation of GRS~1915+105 can be well understood within the current framework
of stellar single and binary evolution.

\acknowledgements We would like to thank Michal Rozyczka, 
Ron Taam, Vicky Kalogera, Ron Webbink, Natasha Ivanova, 
Janusz Zio{\l}kowski and Fred Rasio for comments on this project.  
We acknowledge support from KBN through grant 5P03D01120.

\end{document}